\documentclass[aps,prc,twocolumn,showpacs,floatfix,superscriptaddress,nofootinbib,longbibliography]{revtex4-1}

\usepackage{amsmath}
\usepackage{inputenc}
\usepackage{amssymb}
\usepackage{graphicx}
\usepackage{epsfig}
\usepackage{bm}
\usepackage{color}
\usepackage{float}
\usepackage{dcolumn}
\usepackage{multirow} 
\usepackage[unicode=true,
  linktocpage,
  linkbordercolor={0.5 0.5 1},
  citebordercolor={0.5 1 0.5},
  colorlinks=true,
  linkcolor=blue,
  citecolor=blue,
  urlcolor=blue]{hyperref} 
\usepackage{cgp_macros}
\usepackage{lipsum} 
\usepackage{placeins} 

\graphicspath{{.}}

\newcommand{\beq}{\begin{equation}}
\newcommand{\eeq}{\end{equation}}
\newcommand{\beqa}{\begin{eqnarray}}
\newcommand{\eeqa}{\end{eqnarray}}

\newcommand{\bra}[1]{\left\langle #1 \right|}
\newcommand{\ket}[1]{\left| #1 \right\rangle}
\newcommand{\braket}[2]{\left\langle #1 \vert #2 \right\rangle}

\newcommand{\NNLOsat}{NNLO$_{\rm sat}$}

\begin{document}

\title{Coulomb sum rule  for $^4$He and $^{16}$O from coupled-cluster theory}

\author{J.~E.~Sobczyk}
\affiliation{Institut f\"ur Kernphysik and PRISMA$^+$ Cluster of Excellence, Johannes Gutenberg-Universit\"at, 55128
  Mainz, Germany}

\author{B.~Acharya}
\affiliation{Institut f\"ur Kernphysik and PRISMA$^+$ Cluster of Excellence, Johannes Gutenberg-Universit\"at, 55128
  Mainz, Germany}
  
\author{S.~Bacca}
\affiliation{Institut f\"ur Kernphysik and PRISMA$^+$ Cluster of Excellence, Johannes Gutenberg-Universit\"at, 55128
  Mainz, Germany}
\affiliation{Helmholtz-Institut Mainz, Johannes Gutenberg-Universit\"at Mainz, D-55099 Mainz, Germany}
  
\author{G.~Hagen}
\affiliation{Physics Division, Oak Ridge National Laboratory,
Oak Ridge, TN 37831, USA} 
\affiliation{Department of Physics and Astronomy, University of Tennessee,
Knoxville, TN 37996, USA}

\begin{abstract}
  We demonstrate the capability of coupled-cluster theory to compute
  the Coulomb sum rule for the $^4$He and $^{16}$O nuclei using
  interactions from chiral effective field theory.  We perform several
  checks, including a few-body benchmark for $^4$He. We provide an
  analysis of the center-of-mass contaminations, which we are able to
  safely remove. We then compare with other theoretical results and
  experimental data available in the literature, obtaining a fair
  agreement.  This is a first and necessary step towards initiating a
  program for computing neutrino-nucleus interactions from first
  principles and supporting the experimental long-baseline neutrino
  program with a state-of-the-art theory that can reach medium-mass
  nuclei.

\end{abstract}

\maketitle 

\section{Introduction}

Current neutrino oscillation experiments such as MiniBooNE~\cite{miniboone} and T2K~\cite{T2K} as well as future experiments such as DUNE~\cite{DUNE} and T2HK~\cite{hyperk} are entering a precision phase, with an uncertainty goal of the order of a percent. Accurate extraction of neutrino oscillation parameters from these experiments requires a reliable theoretical treatment of the
scattering of neutrinos with nuclei that constitute the detector material. Presently, the analysis of data is systematically
affected by crude nuclear physics models. Based on existing studies, nuclear structure uncertainties are estimated to be of the order of ten percent~\cite{nustecWP}. A comprehensive understanding of interactions of neutrinos with nuclei built on a cutting edge theory with the capability of reliably estimating uncertainties is urgently needed. Since the energy spectrum of the neutrinos typically ranges from the MeV to the GeV scale, different mechanisms are at play: quasi--elastic scattering, $\Delta$ and $\pi$ production, and deep inelastic scattering. The quasi--elastic process, which dominates the scattering at sub-GeV energies, makes a significant contribution to the total cross section and is the dominant process in the T2HK experiment, in which the neutrino beam peaks at energies of the order of 600 MeV. The quasi--elastic scattering regime is amenable to \textit{ab initio} treatment of nuclei in terms of constituent neutrons and protons, with nuclear interaction and electroweak currents grounded in chiral effective field theory ($\chi$EFT). At present, this approach offers the best opportunity for a direct connection to quantum chromodynamics and a solid uncertainty quantification.

\textit{Ab initio} calculations of neutrino interactions on medium-mass nuclei with modern nuclear
interactions have so far been performed using the Green's function Monte Carlo method~\cite{Lovato2014,Lovato:2015qka,Lovato:2017cux} and the self consistent Green's function method~\cite{Rocco:2018mwt}. By combining the Lorentz integral transform (LIT)~\cite{efros1994,efros2007} with coupled--cluster (CC) theory~\cite{hagen2008} and merging them into a method called LIT-CC, we plan to develop a new \textit{ab initio} approach to  compute general  electroweak responses in medium--mass nuclei. This paper marks a significant milestone in this direction. We build upon the progress made in investigating  
the electron-scattering reactions of light nuclei within the LIT method (see, {\it e.g.}, Refs.~\cite{th4he,BaccaPRL2009,BaccaPRC2009}) and the studies of the giant dipole resonances of light and medium-mass nuclei using the LIT-CC approach~\cite{bacca2013,bacca2014}. Here, we
calculate the Coulomb sum rule (CSR) for $^4$He and $^{16}$O from coupled-cluster theory. While
calculations of the CSR for $^{16}$O from coupled-cluster theory already exist in the literature~\cite{Mihaila}, we present a different approach with a clear goal of extending the LIT-CC formalism to neutrino scattering in the future. We identify, understand and tackle a problem in the LIT-CC method which is  relevant for lepton-nucleus scattering: translationally non-invariant operators induce excitations of the nuclear center-of-mass (CoM) wave function which leads to spuriosities in the calculations of electroweak responses and sum rules. We show that this CoM contamination in the CSR can be removed to a very good approximation by  projecting the final states in the matrix elements out of the subspace spanned by states with CoM excitations. 

The paper is organized as follows. In Sec.~\ref{CSR_theory} we present some general formulas for the Coulomb sum rule. Sec.~\ref{LIT-CC_theory} contains a brief introduction to coupled-cluster theory.
We present several benchmarks in Sec.~\ref{results}, show an analysis of the spurious center-of-mass states in Sec.~\ref{spur} and provide a comparison with other available calculations and experimental data in Sec.~\ref{comp}. Finally, in Sec.~\ref{concl}, we present a brief summary and outlook.

\section{The Coulomb sum rule}
\label{CSR_theory}

The electromagnetic sum rules have provided stringent tests for theories and experiments on electro- and photo-nuclear scattering since the early years of quantum mechanics. The CSR, in particular, has played a major role in building our understanding of nuclear physics beyond a simple model of the nucleus as a collection of non-relativistic point-like protons and neutrons bound by a mean field~\cite{Orlandini_1991}. It continues to guide the development of theories for nuclear electroweak processes (see, {\it e.g.}, Ref.~\cite{lovato2013}) because it allows us to study the response of the nucleus to an external probe induced by the charge operator with the nuclear excitation spectrum integrated out. Formally, the CSR can be defined as an inelastic sum rule~\cite{Orlandini_1991} as
\beqa
\nonumber
\label{csr_a}
m_0^\mathrm{in}(q)&=&\sum_{f_\mathrm{I}}^\infty |\langle f_\mathrm{I}|\rho_\mathrm{I}(q) |0_\mathrm{I} \rangle|^2 - |\langle 0_\mathrm{I}|\rho_\mathrm{I}(q) |0_\mathrm{I} \rangle|^2\,,\\
&=& m_0(q) - Z^2|F_\mathrm{el}(q)|^2
\eeqa
where the first term on the right hand side is the total sum rule of order 0, itself defined as
\beq
m_0(q)=\sum_{f_\mathrm{I}}^\infty |\langle f_\mathrm{I}|\rho_\mathrm{I}(q) |0_\mathrm{I} \rangle|^2 = \langle 0_\mathrm{I}|\rho^{\dagger}_\mathrm{I}(q) \rho_\mathrm{I}(q) |0_\mathrm{I} \rangle
\eeq
 and the term $|F_\mathrm{el}(q)|^2= |\langle 0_\mathrm{I}|\rho_\mathrm{I}(q) |0_\mathrm{I} \rangle|^2/Z^2 $
is the square of the elastic charge form factor of the nucleus. The state $|0_\mathrm{I} \rangle $ is the intrinsic ground state of the nucleus and $|f_\mathrm{I} \rangle $ runs over a complete set of states.

If  we consider point like nucleons and choose the direction of the transfer momentum ${\bf q}$ along the $z$-axis, the charge operator can be written in the intrinsic frame as 
\beq
\rho_\mathrm{I}(q)= \sum_{j=1}^Z e^{i q z'_j}\,,
\eeq
where $z'_j=z_j-Z_\mathrm{CoM}$ are the particle coordinates ($z$-components) relative to the CoM.
In the lab system, however,
the charge operator is 
\beq
\rho(q)= \sum_{j=1}^Z e^{i q z_j}\,,
\label{eq:operator}
\eeq
where $z_j$ are the lab coordinates, as used in coupled-cluster theory.

\subsection{CSR as a ground state expectation value}
\label{csr_method2}

The charge operator written in the lab frame factorizes into the product of an intrinsic operator 
and a CoM operator since 
\beq
\label{rho_factrz}
\rho(q)=\sum_{j=1}^Z e^{i q z'_j} e^{iqZ_\mathrm{CoM}}=\rho_\mathrm{I}(q)~e^{iqZ_\mathrm{CoM}}\,.
\eeq
As a consequence, we have translational invariance for
\beq
{\rho_\mathrm{I}}^\dagger(q) \rho_\mathrm{I}(q)={\rho}^\dagger(q) \rho(q)\,,
\label{ti}
\eeq
at the operatorial level. Furthermore, assuming an exact factorization of the  ground state 
wave functions into intrinsic and CoM part as
\beq
\label{eq:cc_wfn_factorization}
| 0\rangle = | 0_\mathrm{I} \rangle |  0_\mathrm{CoM} \rangle\,,
\eeq
we can formally show that a calculation 
of $m_0(q)$ in the lab frame 
%as performed in coupled-cluster theory 
directly yields the intrinsic-frame result: 
\beqa
\label{csr_lab}
\langle 0|{\rho}^\dagger(q) \rho(q) |0 \rangle &=&\langle 0_\mathrm{I}|{\rho_\mathrm{I}}^\dagger(q) \rho_\mathrm{I}(q) |0_\mathrm{I}  \rangle \langle 0_\mathrm{CoM} | 0_\mathrm{CoM} \rangle \nonumber \\
&=&\langle 0_\mathrm{I}|{\rho_\mathrm{I}}^\dagger(q) \rho_\mathrm{I}(q) |0_\mathrm{I}  \rangle = m_0(q) \,.
\eeqa 

Calculating $m_0(q)$ in this manner as the ground state expectation value of 
the two-body operator ${\rho}^\dagger(q) \rho(q)$ reveals that it is related to the 
Fourier transform of the proton-proton distribution function
\beq
f_{pp}(q)=\frac{1}{Z(Z-1)} \langle 0_\mathrm{I}| \sum_{j\ne k}^Z e^{-i q (z_k-z_j)}| 0_\mathrm{I} \rangle 
\eeq
by~\cite{Orlandini_1991}
\beq
\label{csr}
m_0(q)=Z+Z(Z-1)f_{pp}(q) \,.
\eeq

To obtain the Coulomb sum rule from $m_0(q)$ as in Eq.~(\ref{csr_a}), we need to subtract the elastic part by
calculating the form factor.
In coupled-cluster theory, the form factor is first obtained in the lab frame as
\beq
|F(q)|^2=\frac{1}{Z^2}|\langle 0|\rho(q) |0 \rangle|^2\,.
\label{ff_sq}
\eeq
Assuming again the separable ansatz of the ground state as shown in Eq.~\eqref{eq:cc_wfn_factorization} and using the factorization of the charge operator from Eq.~\eqref{rho_factrz}, this becomes
\beqa
\nonumber
|F(q)|^2 &=&\frac{1}{Z^2}|\langle 0_\mathrm{I}|\rho_\mathrm{I}(q) |0_\mathrm{I} \rangle|^2 |\langle 0_\mathrm{CoM}| e^{iqZ_\mathrm{CoM}}|0_\mathrm{CoM} \rangle|^2\\
&=&|F_\mathrm{el}(q)|^2 |\langle 0_\mathrm{CoM}| e^{iqZ_\mathrm{CoM}}|0_\mathrm{CoM} \rangle|^2\,.
\label{gs}
\eeqa As demonstrated in Refs.~\cite{hagen2009a,jansen2012,hagen2013c}, 
the wave-function factorization in Eq.~\eqref{eq:cc_wfn_factorization}
is valid to a high precision in coupled-cluster theory. Furthermore,
$\ket{0_\mathrm{CoM}}$ is the ground state of a harmonic oscillator
with a fixed frequency $\tilde{\omega}$ that can be determined by
requiring that the coupled-cluster ground state expectation value of
the CoM Hamiltonian,
$P_\mathrm{CoM}^2/(2M)+1/2 M \tilde{\omega}^2
R_\mathrm{CoM}^2-3/2\hbar\tilde\omega$, vanishes. One can therefore
easily calculate
$|\langle 0_\mathrm{CoM}| e^{iqZ_\mathrm{CoM}}|0_\mathrm{CoM}
\rangle|^2$.  $|F_\mathrm{el}(q)|^2$ thus obtained from
Eqs.~\eqref{ff_sq} and \eqref{gs} is then subtracted from $m_0(q)$
given by Eq.~\eqref{csr_lab} yielding the Coulomb sum rule
$m_0^\mathrm{in}(q)$. Calculating the CSR in this manner is denoted as
``method A'' in the rest of the paper. We expect this approach to be
free of contaminations from the CoM wave function to a good level of
approximation. For $^4$He, we will numerically verify this by
comparing calculations performed in the lab frame using
coupled-cluster theory with calculations obtained from hyperspherical
harmonics working in the intrinsic frame.

\subsection{CSR as sum over multipoles}
\label{csr_method1}

The approach of Sec.~\ref{csr_method2} is not applicable in general to the calculation of sum rules and response functions of the large set of electroweak charge and current operators we will need to include in neutrino scattering calculations. The standard approach while working with angular momentum eigenstates is to perform a multipole decomposition of the electroweak operators, see, {\it e.g.}, Ref.~\cite{bijaya}. We therefore expand the charge operator into multipoles as
\beq
\rho(q)=\sum_{J=0}^{\infty} [\rho(q)]^J\,.
\label{mult_rho}
\eeq
In practice, the infinite sum over $J$ needs to be truncated at a finite multipolarity. 
The number of multipoles needed to achieve convergence depends on the momentum $q$ and the size of the nucleus. We obtain $m_0(q)$ as a coherent sum,  
\beq
m_0(q) = \sum_{J=0}^{\infty} m_0^J(q)\,,
\eeq
where 
\beqa
\label{mult_str}
m_0^J(q) &=& \langle 0|[\rho^\dagger(q)]^J [\rho(q)]^J |0  \rangle \nonumber \\
&=& \sum_{f^J}  \langle 0|[\rho^\dagger (q)]^J| f^J \rangle \langle f^J |[\rho (q)]^J |0  \rangle\,
\eeqa
is the total strength of each multipole operator.
Here we have inserted the completeness relation in terms of states $|f^J\rangle$ labeled by 
total angular momentum quantum number $J$, only retained non-vanishing terms, 
and have restricted ourselves to 
the case where the nuclear ground state $|0  \rangle$ has zero total angular momentum. 
The elastic part can be subtracted by simply restricting the sum over $f^J$ to $|f^J\rangle\neq|0\rangle$ 
in Eq.~\eqref{mult_str}. 
In this way we remove the contribution coming from the lab form factor
$|F(q)|^2$, which is not the same as the removal of
$|F_\mathrm{el}(q)|^2$ in method A.

At this point it is important to note that, since the operator
$[\rho (q)]^J$ is not translationally invariant, the states
$| f^J \rangle$ can contain CoM excitations. Translationally
non-invariant operators can generate CoM
excitations by acting on $|0_\mathrm{CoM}\rangle$ even if the ground
state $|0\rangle$ factorizes exactly as shown in
Eq.~\eqref{eq:cc_wfn_factorization}.  We expect such spurious
excitations to make significant contributions to the sum rule for the
lowest multipoles. In Sec.~\ref{spur}, we will discuss how we remove
these CoM excitations and demonstrate it with a practical numerical
implementation.
Let us note that spurious CoM contamination could also be present when
using translationally invariant operators such as the electric
dipole~\cite{bacca2014} if the factorization shown in
Eq.~\eqref{eq:cc_wfn_factorization} is inexact. In the LIT-CC method,
they can be removed in the Lanczos algorithm, as done in
Ref.~\cite{bacca2014}. In this work we will use a similar technique
(described below), and calculating the CSR in this way will be called
``method B'' in the rest of the paper.

\section{Coupled-cluster theory}
\label{LIT-CC_theory}

In coupled-cluster
theory~\cite{coester1958,coester1960,kuemmel1978,mihaila2000b,dean2004,wloch2005,hagen2010b,binder2013b,hagen2014}
one uses  the similarity transformed Hamiltonian,
\begin{equation} 
\overline{H}_N= e^{-T} H_N e^T, \;\; T = T_1 + T_2 + \ldots,
\end{equation} 
where $H_N$ is normal-ordered with respect to a single-reference state
$\vert \Phi_0\rangle$ (usually a Slater determinant obtained from a Hartree-Fock calculation), and $T$ is an
expansion in particle-hole excitations with respect to this
reference. 
 The operator $T$ is typically truncated at some low rank particle-hole
excitation level.

Because the similarity transformed Hamiltonian is non-Hermitian, one has to
 compute  both the left and right eigenstates in order to
evaluate expectation values and transition strengths. The right ground state is $\ket{0}=\ket{\Phi_0}$
while left ground state is
obtained as 
\begin{equation}
  \langle 0 \vert =\langle \Phi_0 \vert (1+\Lambda), \;\; \Lambda = \Lambda_1 + \Lambda_2 + \ldots,
\end{equation}
where $\Lambda $ is a sum of particle-hole de-excitation
operators.

In the LIT-CC method, we use the coupled-cluster
equation-of-motion (EOM) method~\cite{stanton1993,hagen2013c} together with a
non-symmetric Lanczos algorithm to solve for a generalized
non-Hermitian eigenvalue problem with a source term (see
Ref.~\cite{miorelli2018} for details).
In practice one deals with the excited states of the Hamiltonian defined as
\begin{eqnarray} 
\label{eq:eomcc}
\nonumber
\overline{H}_N R_\mu \vert\Phi_0 \rangle = E_\mu R_\mu \vert \Phi_0
\rangle\, ,  \\
\langle \Phi_0 \vert L_\mu \overline{H}_N = E_\mu \langle \Phi_0\vert
L_\mu \, ,
\end{eqnarray}
where $R_\mu $ and $L_\mu $ are linear expansions in
particle-hole excitations as well.
To compute an electromagnetic transition strength from the ground to an excited
state in coupled-cluster theory, one performs the following calculation (see Ref.~\cite{miorelli2018}) 
\begin{eqnarray} 
 \vert \langle f_{\mu} \vert \hat{\Theta}\vert 0 \rangle \vert^2 &  =  & \langle 0 \vert \hat{\Theta}^\dagger\vert f_{\mu} \rangle \langle f_{\mu} \vert \hat{\Theta}\vert 0 \rangle =\\
&=&
\nonumber
\langle \Phi_0 \vert (1 + \Lambda ) \overline{\Theta}_N^\dagger R_\mu \vert \Phi_0 \rangle 
\langle \Phi_0 \vert L_\mu \overline{\Theta}_N \vert \Phi_0 \rangle \,,
\end{eqnarray}
where $\overline{\Theta}_N \equiv e^{-T}\Theta_Ne^T$ is the similarity 
transformed normal-ordered operator. In this work the  operator is taken
to be an electromagnetic Coulomb multipole of rank $J$, namely  $\Theta=[\rho(q)]^J$ (see also Ref.~\cite{bijaya}), 
and a summation on the multipoles is performed afterwards.
The Coulomb multipoles are one-body
operators, so that the Baker-Campbell-Hausdorff expansion for $\overline{\Theta}_N$
terminates at doubly nested commutators
\begin{equation}
  \overline{\Theta}_N= \Theta_N + \left[ \Theta_N,T\right] + {1\over 2}\left[ \left[ \Theta_N,T\right],T\right]\,.
  \label{bhc}
\end{equation}
Finally, the total multipole strength as calculated in method B is~\cite{miorelli2018}
\begin{eqnarray}
\label{eq:noresp} 
m_0^J(q) & = & \langle \nu_L^J \vert \nu_R^J \rangle = \langle \Phi_0 \vert
(1+\Lambda) \overline{\Theta}_N^\dag \cdot \overline{\Theta}_N \vert \Phi_0 \rangle \,,
\end{eqnarray}
where we used the closure relation. Let us label the spurious
excitations of angular momentum $J$ by $s^J$. The spurious states as well as the
ground state are obtained from the coupled-cluster EOM technique using
an iterative Arnoldi algorithm. Having obtained the converged left
$\langle{s}_L^J\vert$ and right $\vert {s}_R^J\rangle$ spurious states we
project them out of Eq.~\eqref{eq:noresp}  by multiplying the left $\langle \nu_L^J  \vert  $ and right
$\vert \nu_R^J \rangle $ states with $\left( {\bf 1} - \vert s_R^J \rangle
\langle s_L^J\vert \right)$ to obtain
\begin{equation}
\label{eq:m_J_bar} 
\bar{m}_0^J(q) = \langle \nu_L^J \vert \nu_R^J \rangle-\langle \nu_L^J \vert s_R^J \rangle \langle s_L^J\vert \nu_R^J \rangle,
\end{equation}
(see Sec.~\ref{spur} below for further details).

In both methods A and B, the computation of the CSR is performed in
the so-called coupled-cluster with singles-and-doubles (CCSD) approximation,
where all the particle-hole expansions mentioned above are truncated
at the two-particle-two-hole level.  While we have developed the
necessary technology to deal with leading-order triple corrections, in
Ref.~\cite{miorelli2018} we found that they are negligible for
non-energy weighted sum rules (as the CSR is), furthermore these
excitations increase the computational cost by about an order of
magnitude. Hence, we neglect them at this point and focus on studying
the momentum dependence of our results.

\section{Benchmarks}
\label{results}
We employ Hamiltonians from $\chi$EFT in our calculations.
Specifically, we employ two interactions. First, we use the
nucleon-nucleon ($NN$) chiral potential by Entem and
Machleidt~\cite{entem2003} (labeled by N$^3$LO-EM) without
supplementing it with a three-nucleon (3N) interaction.  This will
serve to perform tests with hyperspherical harmonics
expansion~\cite{EIHH1} in $^4$He and to perform an analysis of the
center of mass spuriosities.  Second, we use the \NNLOsat~
interaction~\cite{ekstrom2015}, that includes both $NN$ and $3N$
interactions at next-to-next-to leading order in $\chi$EFT. In this
case, the parameters entering the $NN$ and $3N$ interactions were
adjusted to $NN$ phase shifts and to energies and charge
radii of selected nuclei up to $^{24}$O. We use this interaction as it
has been shown to work well for the nuclei studied in this paper.

\begin{figure}[hbt]
  \includegraphics[width=0.5\textwidth]{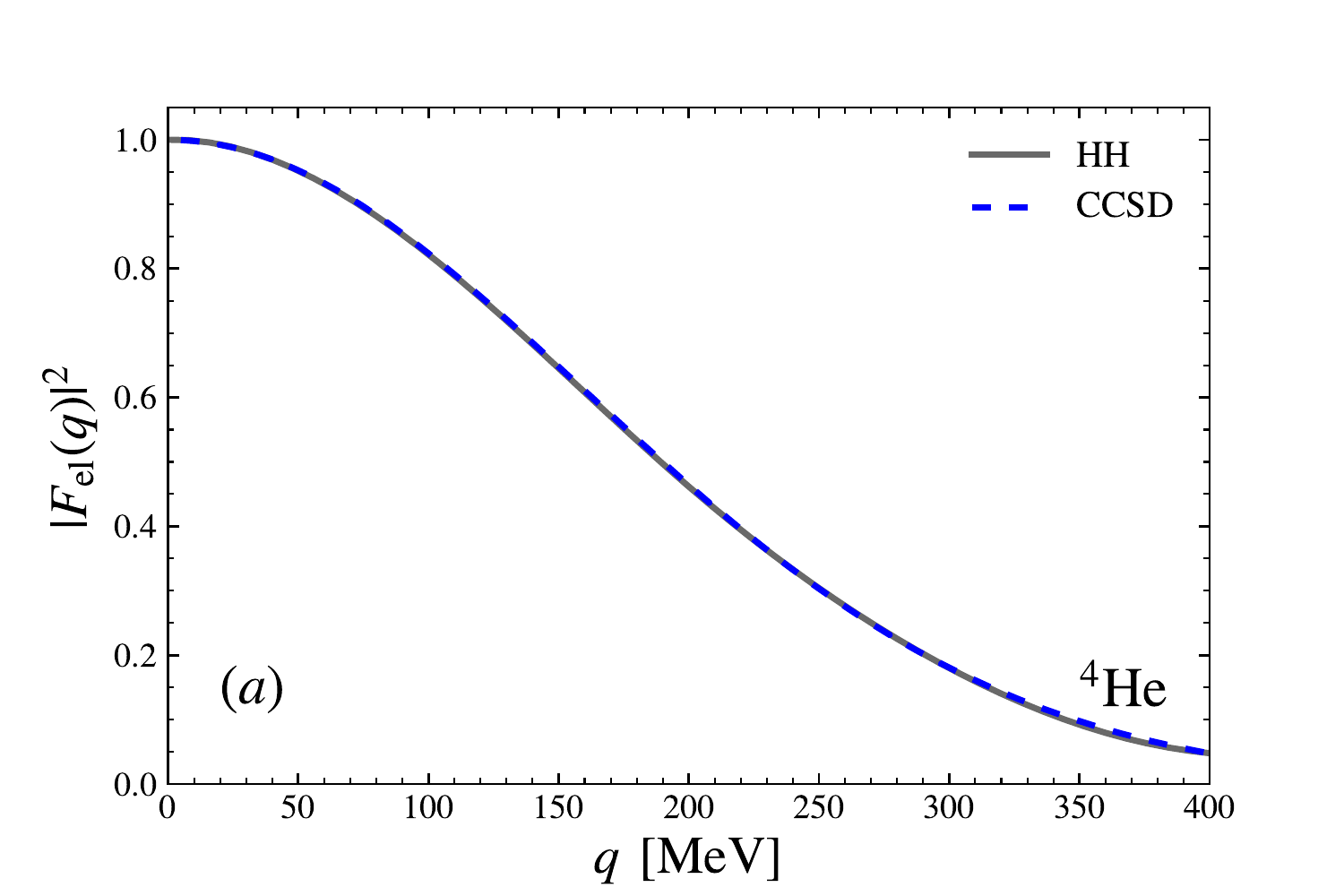}
  \includegraphics[width=0.5\textwidth]{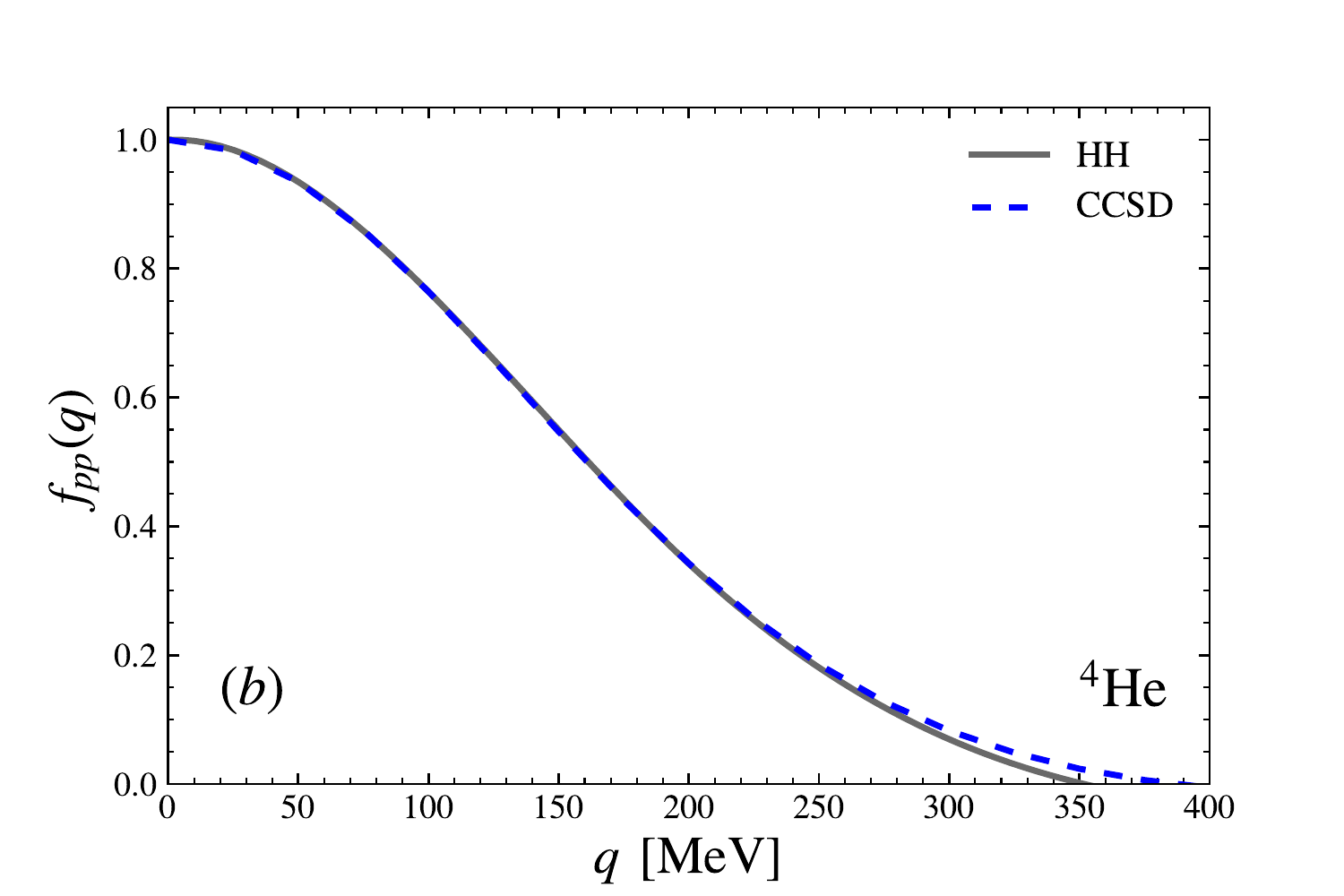}
  \caption{Coupled-cluster calculations in CCSD compared against HH using the N$^3$LO-EM interaction. 
  The two separate components making up the CSR, namely the squared elastic form factor $|F_\mathrm{el}(q)|^2$  (panel $(a)$) and  $f_{pp}(q)$ (panel $(b)$) are shown  as functions of the momentum transfer $q$.}
  \label{bench}
\end{figure}

{\it Few-body benchmark for $^4$He---} We start presenting our results
from coupled-cluster theory obtained from method A using the
N$^3$LO-EM $NN$ interaction. In Fig.~\ref{bench}, we show separately
the elastic squared form factor $|F_\mathrm{el}(q)|^2$ and the
proton-proton distribution function $f_{pp}(q)$, that are related to
CSR by Eqs.~\eqref{csr_a} and \eqref{csr}. The calculations is
performed in the CCSD scheme for a model space of $15$ major
oscillator shells and an underlying harmonic oscillator frequency of
$\hbar \omega = 20$ MeV. In order to remove the CoM contamination from
$|F_\mathrm{el}(q)|^2$ (see Eq.~\eqref{gs}), an underlying CoM
frequency $\hbar \tilde\omega = 20$ MeV was employed, which is known
from Ref.~\cite{hagen2009a}. We compare it to calculations performed
with the hyperspherical harmonics method (HH)~\cite{barnea2001} using
the same potential.  Such few-body computations have been performed
with a model space of $K_{max}=16$ in the HH method, where the
accuracy is expected to be of about a percent. As one can see, the
agreement is very good over a wide range of momentum transfer $q$,
indicating that for the $^4$He and chiral potentials triples and
quadrupole excitations missing in the CCSD approximation can safely be
neglected. Only for the function $f_{pp}(q)$, one can see a deviation
of the two curves at the largest shown momenta. However, since this
quantity is much smaller than 1 at high
$q$ values, the observed alignment of the HH and CCSD curves leads to
a nice agreement at the CSR level. This allows us to infer that the
factorization of the wave function into a CoM and an intrinsic part
must hold to a large extent, since a calculation performed in the lab
frame agrees with the one in intrinsic frame.

{\it Comparison of two calculational methods---}
At this point, we proceed with tests of the coupled-cluster computations, namely we want to verify if method A and B agree with each other.
\begin{figure}[hbt]
  \includegraphics[width=0.5\textwidth]{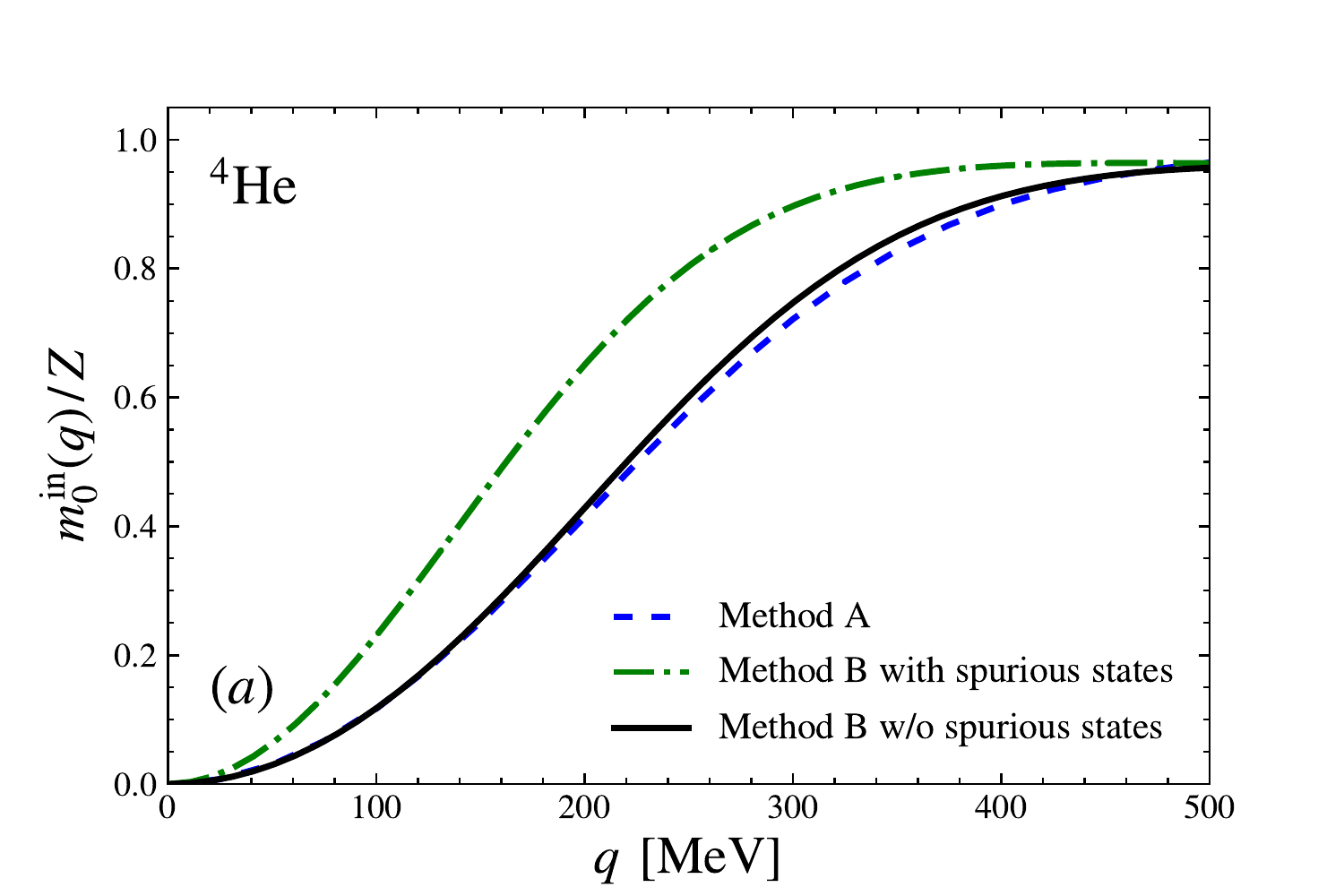}
  \includegraphics[width=0.5\textwidth]{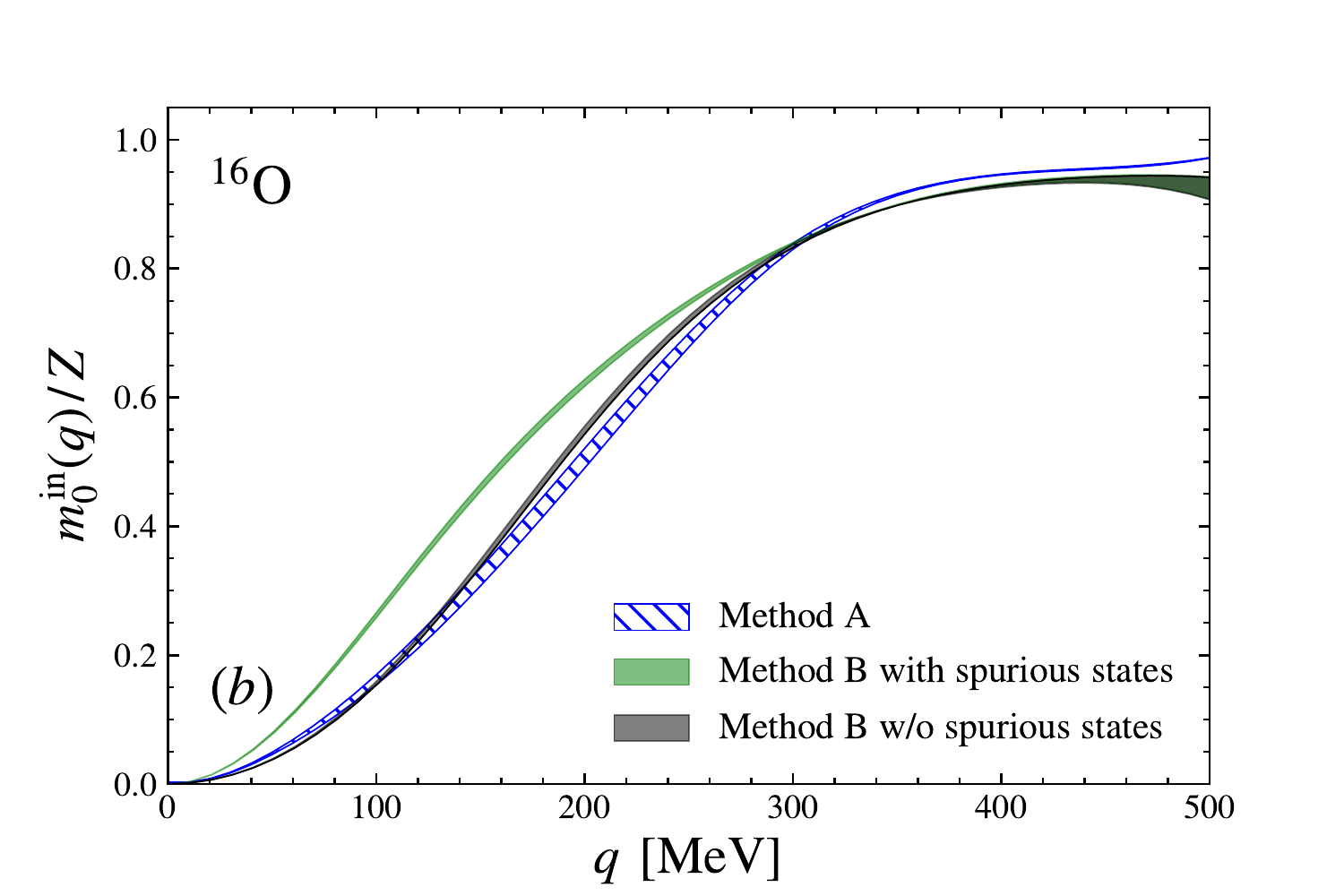}
  \caption{Comparison of method A and B: for $^4$He (panel ($a$)) and $^{16}$O (panel ($b$)) using the 
 N$^3$LO-EM interaction. For method B we show the case with  and without (w/o) spurious CoM states. }
  \label{methods}
\end{figure}
As mentioned above, method B may include spurious states that need to
be removed.  Because identifying all spurious states by running the
full protocol described in Refs.~\cite{jansen2012,parzuchowski2017} is
computationally demanding, we adopt a more pragmatic strategy in this
work.  For $^4$He, we omit from the Lanczos iterations all states with
excitation energy below 15 MeV, as they have to be spurious given that
$^4$He has no excited state below the proton emission threshold.  The
omitted states were a $1^-$ state approximately at energy
$5\times10^{-2}$~MeV above the ground state and a $2^+$ state at
energy 10~MeV above the ground state. For $^{16}$O, we remove a $1^-$
state at $0.15$~MeV, which also must be spurious as this nucleus does
not have any excited states at such low energy.

In Fig.~\ref{methods} we show a comparison between method A and method
B for $^4$He and $^{16}$O using the N$^3$LO-EM interaction. For method
B we show the curve obtained in case we do not remove the spurious
states in the Lanczos procedure as well as the one where we remove
such states.  Calculations are performed for a model space of 15
oscillator shells and a harmonic oscillator frequency of
$\hbar \omega = 20$ MeV for all the three curves. In case of $^{16}$O
we show a band obtained by the difference between the model spaces
with 15 and 13 major shells. This gives us an idea of the uncertainty in a
model space size that will be achievable in other calculations where
we will add $3N$ interactions.

First of all, we see that the removal of the spuriosities has a much
larger effect in $^4$He than in $^{16}$O as expected, since the
heavier the nucleus, the smaller the CoM contamination must
be. Second, we see that the results of method B w/o spurious states
agree quite nicely with those of method A for both nuclei.
\begin{figure}[hbt]
  \includegraphics[width=0.5\textwidth]{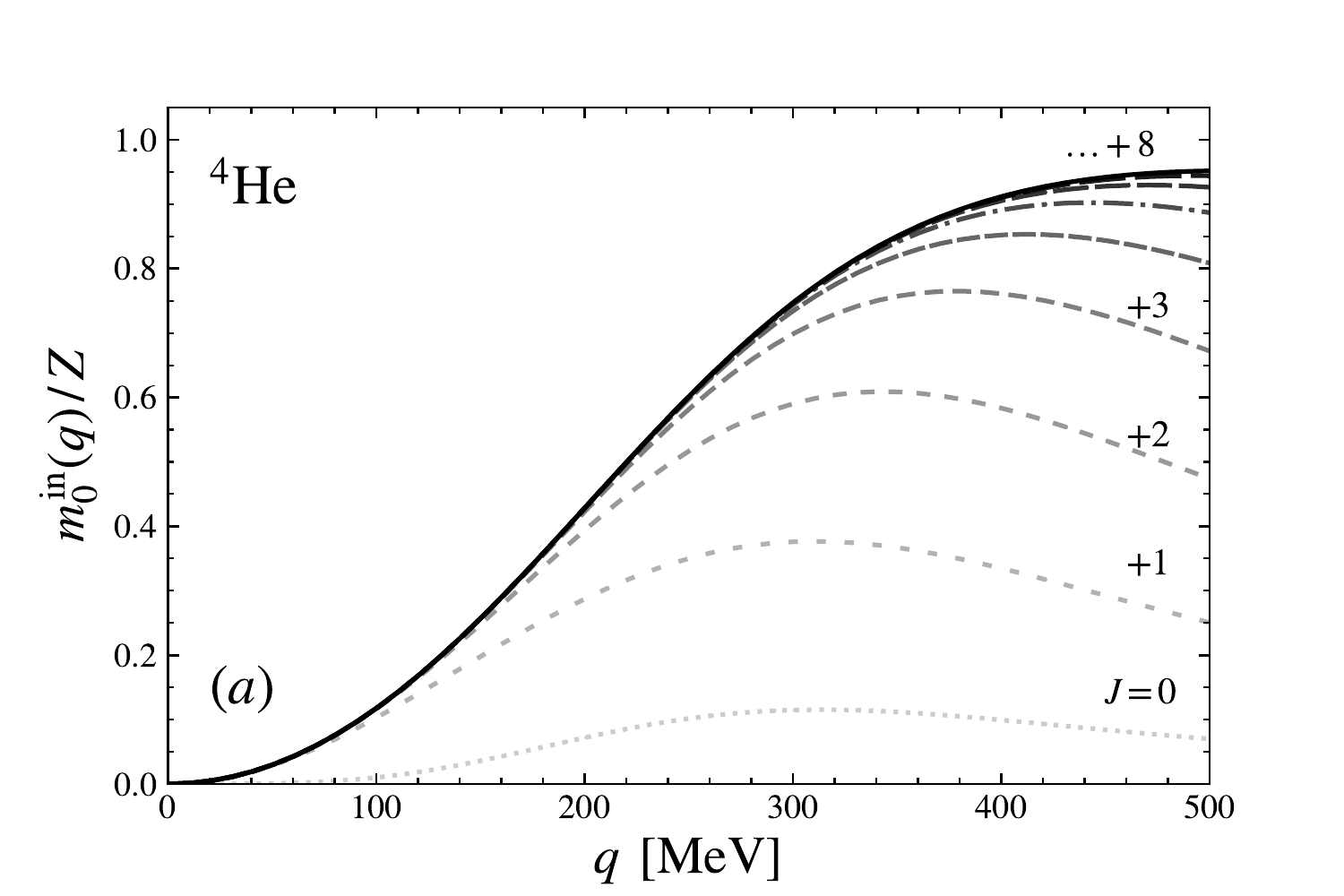}
  \includegraphics[width=0.5\textwidth]{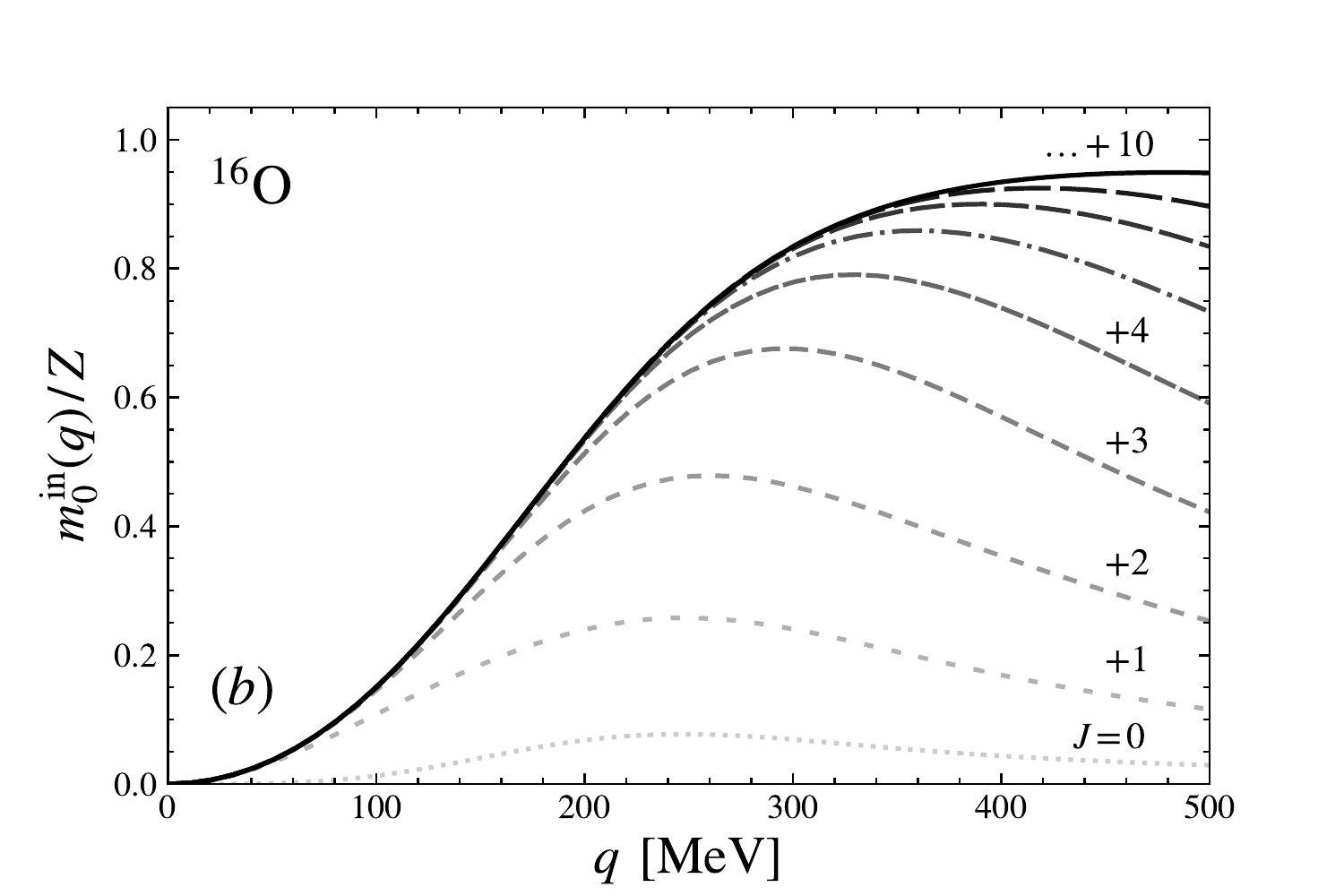}
  \caption{Cumulative sums of the multipole expansion for the Coulomb sum rule of $^4$He (panel ($a$)) and $^{16}$O (panel ($b$)) using the  N$^3$LO-EM interaction in method B w/o spurious states.}
  \label{mult}
\end{figure}
The removal of spuriosities brings the agreement at $q=100$ MeV from $50\%(38\%)$ to $0.5\%(3.5\%)$ for $^4$He($^{16}$O), while at $q=200$ MeV, it brings the agreement from $35\%(20\%)$ to $3\%(8\%)$.
We consider this comparison satisfactory and from now on we will use method B only w/o the spurious states to perform further analysis.

{\it Convergence of multipole expansion---}
A natural question in method B is how many multipoles are needed in
the expansion of Eq.~(\ref{mult}). We show this for the N$^3$LO-EM
$NN$ interaction in Fig.~\ref{mult}.  These calculations are performed for
the same model space as above. Clearly, the convergence in terms of
multipoles strongly depends on the momentum transfer, as expected. 
Furthermore, we see
that the convergence is faster for the smaller $^4$He nucleus than for
the larger $^{16}$O nucleus. 
For $q$ below 200 MeV only four multipoles are needed, while for $q=500$
MeV nine and eleven multipoles are needed to reach satisfactory
convergence for $^4$He and $^{16}$O, respectively. The information on the number of needed
multipoles will be relevant to estimate the amount of computing time
requested to compute response functions with the LIT-CC
method. Otherwise, the CSR can be most efficiently computed with
method A instead.

\section{Analysis of spurious states}
\label{spur}

In method B we find that by far the largest spurious contribution that
we remove, comes from the $1^-$ state.  In order to understand the
nature of the spurious states we removed, let us assume for the moment
that the factorization of $| 0 \rangle$ into
$| 0_\mathrm{I} \rangle | 0_\mathrm{CoM}\rangle$ is exact, with
$\ket{0_\mathrm{CoM}}$ being the ground state of a quantum harmonic
oscillator.  The non-spurious excitations are those that have the form
$| f_\mathrm{I}^J \rangle | 0_\mathrm{CoM}\rangle$, where
$| f_\mathrm{I}^J \rangle$ are excitations of the intrinsic wave
function, labeled by total angular momentum quantum number $J$.  The
simplest possible form of the spurious states is then
$ | 0_\mathrm{I} \rangle | f^J_\mathrm{CoM} \rangle$, where
$| f^J_\mathrm{CoM} \rangle$ are excitations of the CoM wave function.
The completeness relation used in Eq.~\eqref{mult_str} then separates
into the projectors onto the coupled-cluster ground state, the
subspace spanned by non-spurious excitations and the subspace spanned
by spurious excitations, as 
\beqa
\sum_{f^{J}} | f^{J} \rangle \langle f^{J} | & \simeq & | 0_\mathrm{I} \rangle | 0_\mathrm{CoM} \rangle \langle 0_\mathrm{CoM} | \langle 0_\mathrm{I} | \nonumber\\
&+& \sum_{f_\mathrm{I}^{J}} | f_\mathrm{I}^{J} \rangle | 0_\mathrm{CoM} \rangle \langle 0_\mathrm{CoM} | \langle f^{J}_\mathrm{I} | \nonumber\\
&+& \sum_{f^{J}_\mathrm{CoM}}| 0_\mathrm{I} \rangle | f_\mathrm{CoM}^J
\rangle \langle f_\mathrm{CoM}^J | \langle 0_\mathrm{I} |\,.
\label{eq:sp_st_structure}
\eeqa
Inserting this relation in Eq.~\eqref{mult_str}, we obtain 
\beq
m_0^J(q) =Z^2 |F(q)|^2 \, \delta_{J0}  +  [m_0^\mathrm{in}(q)]^J + [m_0^\mathrm{sp}(q)]^J\,,
\eeq
where the first term on the right hand side is the squared elastic form factor in the lab frame and only 
contributes to the monopole term, 
the second term is the multipole strength of the CSR and the third term is the contribution of the 
spurious states to $m_0^{J}(q)$. 

The Coulomb sum rule,  
\beq
m_0^\mathrm{in}(q) = \sum_{J=0}^\infty [m_0^\mathrm{in}(q)]^J\,,
\eeq
which is calculated in practice by adding a finite number of multipoles until convergence is achieved, 
is then free of contributions from CoM excitations. The spurious state contribution $[m_0^\mathrm{sp}(q)]^J$ is given by the squared matrix element
\beqa
[m_0^\mathrm{sp}(q)]^J &&= \vert \langle 0_\mathrm{I}|\langle {f_\mathrm{CoM}^J}  |[\rho]^J_0 |0  \rangle \vert^2\nonumber\\ 
&&=
\vert \langle 0_\mathrm{I}|\langle {f_\mathrm{CoM}^J} |\left[[\rho']^{J_1}_{m_1} \otimes [e^{iq Z_\mathrm{CoM}}]^{J_2}_{m_2}  \right]^J_0 |0_\mathrm{I} \rangle |  0_\mathrm{CoM} \rangle \vert^2\nonumber\\
&&=\vert\sum_{m} C^{J 0}_{0 0 J 0} 
\langle 0_\mathrm{I}|[\rho']^{0}_{0} |0_\mathrm{I} \rangle  \langle f_\mathrm{CoM}^J | [e^{iq Z_\mathrm{CoM}}]^{J}_{0}  |  0_\mathrm{CoM} \rangle\vert^2 \nonumber\\
&& =Z^2|F_\mathrm{el}(q)|^2 K_\mathrm{CoM}^J(q)\,,
\eeqa
where $C_{J_1 m_1J_2m_2}^{J m}$ are Clebsch-Gordan coefficients and  
\beq
\label{eq:k_function}
K_\mathrm{CoM}^J(q)=|\langle f_\mathrm{CoM}^J | [e^{iq Z_\mathrm{CoM}}]^{J}_{0}  |  0_\mathrm{CoM} \rangle|^2
\eeq
depend only on the CoM.

It was shown in Ref.~\cite{hagen2009a} that the CoM part of the coupled-cluster ground state, $|0_\mathrm{CoM} \rangle$, is well approximated by the ground state of a three-dimensional harmonic oscillator Hamiltonian with energy eigenvalues $E= \hbar \tilde{\omega} (N + \frac{3}{2})$, where the number of oscillator quanta $N$ is related to the radial quantum number $N_r$ and the angular momentum quantum number $J$ by $N=2 N_r + J$ with $J \leq N$. The oscillator length parameter is $b=\sqrt{\frac{\hbar}{2 M \tilde{\omega}}}$, where $M$ is the mass of the nucleus.

 Here, we make the ansatz that the CoM part of the spurious state, $| f_\mathrm{CoM}^J \rangle$, is given by the excited oscillator state $|\Psi_{0J0}^{\rm CoM} \rangle$. 
The CoM functions $K_\mathrm{CoM}^J(q)$ can then be written as 
\beq
\label{k_functions_sum}
K_\mathrm{CoM}^J(q)= \vert \langle \Psi_{0J0}^{\rm CoM} | [e^{iq Z_\mathrm{CoM}}]^{J}_{0}  |  \Psi_{000}^{\rm CoM} \rangle\vert^2\,,
\eeq
and calculated using co-ordinate representations of the oscillator states,
\begin{align}
 \langle \mathbf{R}_\mathrm{CoM} \vert  \Psi_{N_rJM}^{\rm CoM} \rangle & = \sqrt{\frac{2N_r !}{b^{3}
\Gamma\left(N_r +J+3/2\right)}}\left(\frac{R_{\rm CoM}}{b}\right)^{J} \nonumber\\ 
& \quad \times \exp{\left(-\frac{R_{\rm CoM}^{2}}{2b^{2}}\right)} 
 L_{N_r}^{J+\frac{1}{2}} \left(\frac{R_{\rm CoM}^{2}}{b^{2}}\right) \nonumber\\
& \qquad Y_{JM} (\hat{\mathbf{R}}_{\rm CoM})\,,
\end{align}
where $L_n^\ell(x)$ are associated Laguerre polynomials which satisfy $L_0^\ell(x) = 1$ for all values of 
$\ell$ and $x$. This yields
\begin{equation}
\label{k_functions_simple}
K_\mathrm{CoM}^J(q) = \frac{\sqrt{\pi} (2J+1)}{2 \Gamma (J + 3/2)} \,\left(\frac{b q}{2} \right)^{2J} 
\exp\left(-\frac{b^{2}q^{2}}{2}\right)\,.
\end{equation}

A closed-form expression for $K_\mathrm{CoM}^J(q)$ can also be
obtained for a more general ansatz,
$| f_\mathrm{CoM}^J \rangle = |\Psi_{NrJ0}^{\rm CoM} \rangle$, which
allows the spurious states to contain radial excitations with
$N_r>0$. This derivation, along with a detailed numerical analysis of
such excitations, will be communicated through a separate
publication~\cite{BB}. However, we do find that radial excitations are
negligible in the following analysis, hence we keep our ansatz as
simple as possible.

Let $\bar m_0^{J}(q)$ be the multipole strength calculated by removing
a spurious state of angular momentum $J$ (and the ground state for $J=0$) in the Lanczos procedure, defined in Eq.~\eqref{eq:m_J_bar}. 
The difference $m_0^{J}(q) - \bar{m}_0^{J}(q)$ can then be obtained
 in the LIT-CC method which allows us to numerically check
whether \beq
\label{ansatz}
m_0^{J}(q) - \bar{m}_0^{J}(q)  =  Z^2|F_\mathrm{el}(q)|^2  K_\mathrm{CoM}^J(q) \,,
\eeq
holds for the expression for $K_\mathrm{CoM}^J(q)$ given by Eq.~\eqref{k_functions_simple}. We thus validate all our assumptions about the nature of the CoM spuriosity.

It is important to realize that summing Eq.~\eqref{ansatz} over $J$ we obtain\footnote{This would be exactly equal to $Z^2|F_\mathrm{el}(q)|^2$ if radial excitations were included~\cite{BB}.}
\beq
\label{ansatz-sum}
m_0(q) - \bar{m}_0(q) =Z^2 |F_\mathrm{el}(q)|^2 \sum_{J=0}^\infty
K_\mathrm{CoM}^J(q) \approx Z^2|F_\mathrm{el}(q)|^2\,.  \eeq On the left
hand side we have the sum of all the spurious states (with the ground
state in the lab frame included) obtained numerically. It should
correspond to the elastic form factor squared, so one can see here a
direct connection with Eq.~\eqref{csr_a}.
\begin{figure}[hbt]
  \includegraphics[width=0.5\textwidth]{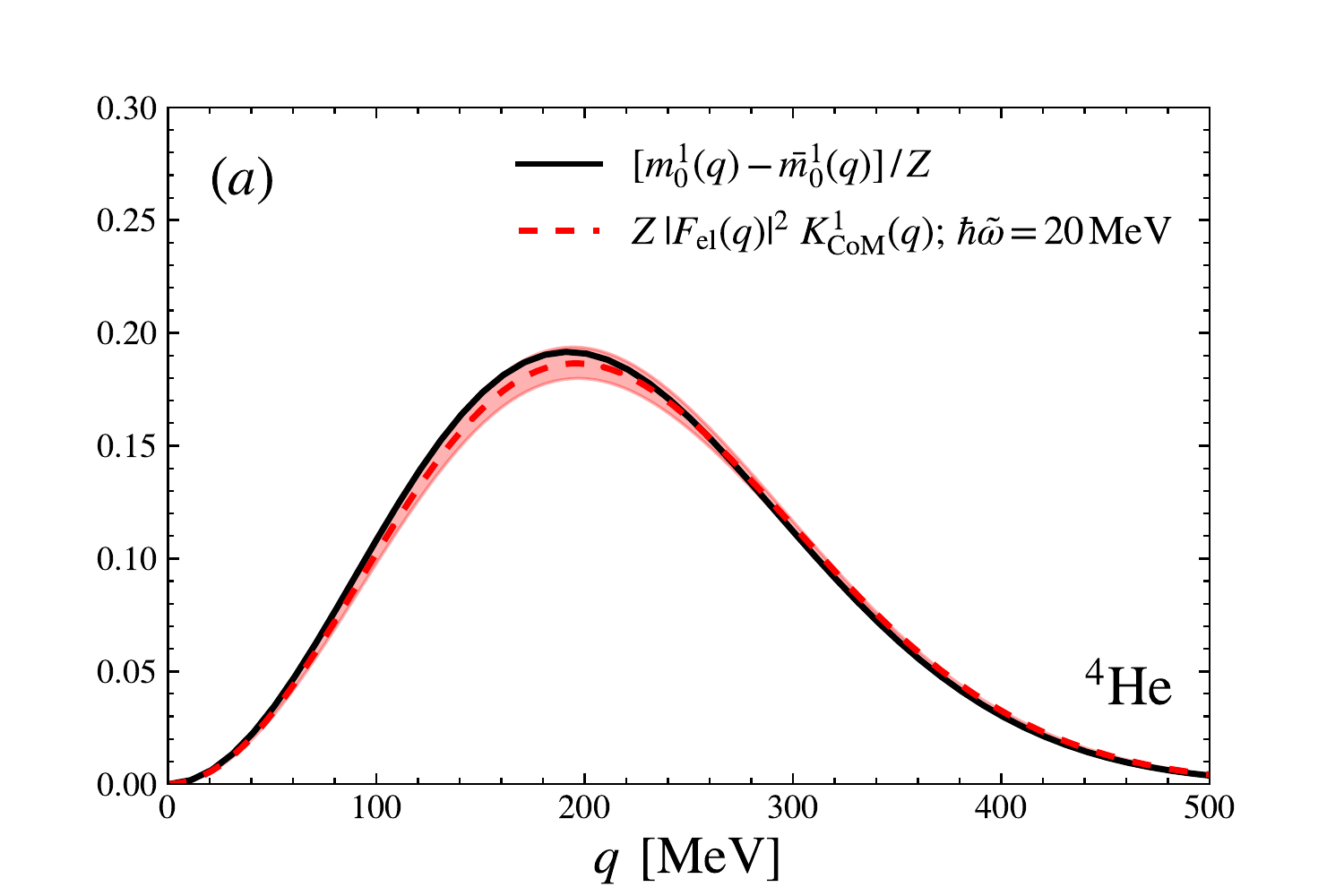}
  \includegraphics[width=0.5\textwidth]{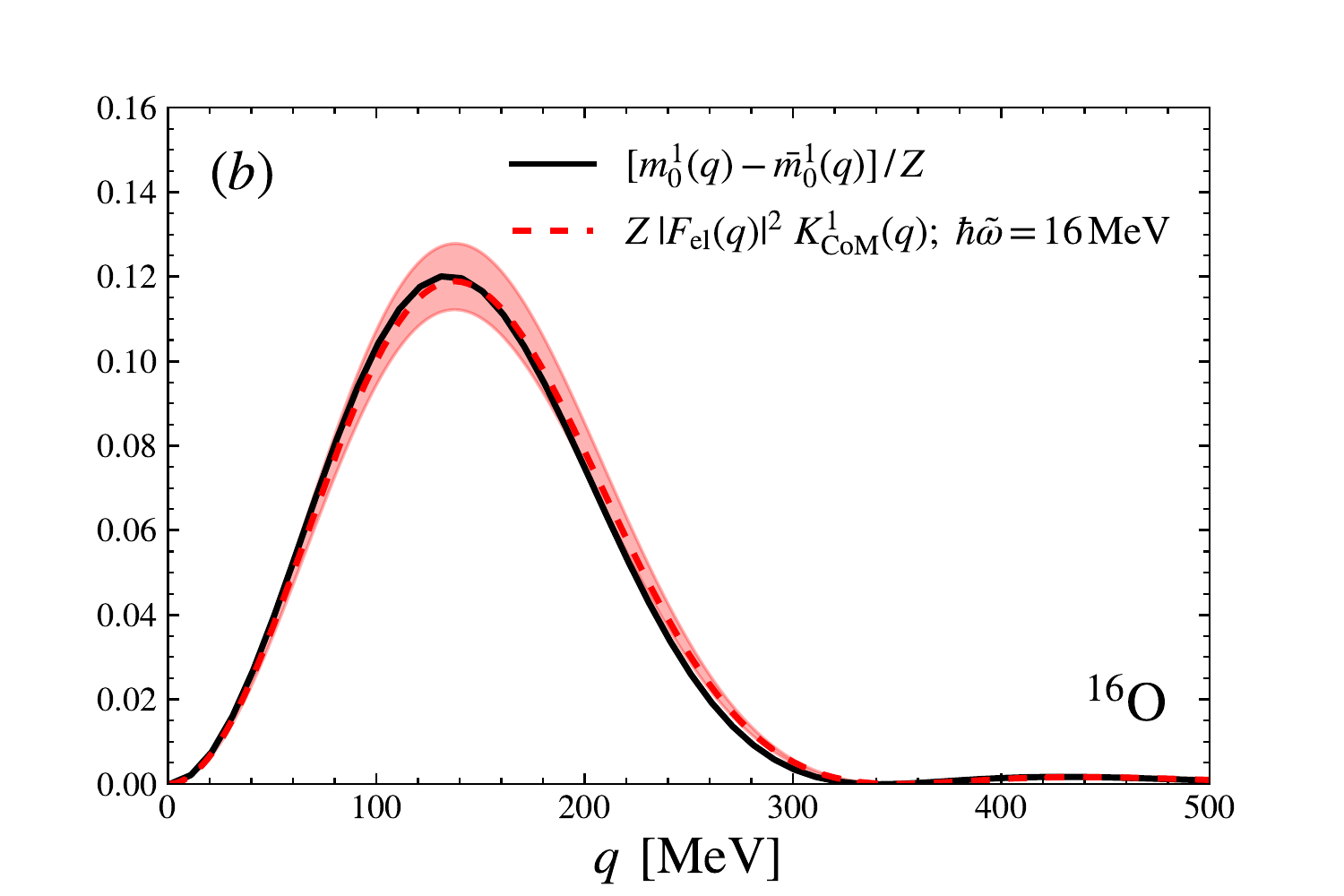}
  \caption{Spurious states $1^-$ for $^4$He (panel ($a$)) and $^{16}$O (panel ($b$)) compared with an ansatz for the CoM function from Eq.~\eqref{k_functions_simple}. The band for $^4$He is obtained with $\hbar \tilde \omega = 20\pm1$ MeV. For $^{16}$O the band corresponds to $\hbar \tilde \omega = 16\pm1$ MeV.}
  \label{fig:k_functions}
\end{figure}

Let us note that for $J=0$, we have that $K_\mathrm{CoM}^0(q)=|\langle 0_\mathrm{CoM}| e^{iqZ_\mathrm{CoM}}|0_\mathrm{CoM} \rangle|^2$, and so in Eq.~\eqref{ansatz} on the right hand side we obtain $Z^2|F(q)|^2$ which stays in agreement with Eq.~\eqref{gs}. This way we show that Eq.~\eqref{ansatz} holds for $J=0$.

In Fig.~\ref{fig:k_functions}, we plot the contributions of the $1^-$
spurious states to the sum rules for $^4$He and $^{16}$O, given
numerically in the LIT-CC theory by the difference
$m_0^{1}(q)-\bar{m}_0^{1}(q)$, along with their $K_\mathrm{CoM}^1(q)$
functions given by Eq.~\eqref{k_functions_simple} and obtain an
excellent agreement. For the employed interaction N$^3$LO-EM, it was
shown in Ref.~\cite{hagen2009a} that the frequency
$\hbar\tilde{\omega}$ of $\vert 0_{\mathrm CoM} \rangle$ is close to
19~MeV for $^4$He and 16~MeV for $^{16}$O in the $\Lambda-$CCSD(T)
approximation, respectively. We therefore varied the frequencies
$\hbar\tilde{\omega}$ in a range of $\pm~1$~MeV around these values.
The agreement shown in Fig.~\ref{fig:k_functions} can only be obtained
for this narrow range for which the ground state approximately
factorizes into intrinsic and CoM wave functions, further supporting
our ansatz for the form of $| f_\mathrm{CoM}^J \rangle\,$.  We would
like to emphasize here that the frequency $\hbar \tilde \omega=20$ MeV
for $^4$He coincides with the value employed to obtain
$F_\mathrm{el}(q)$, see Fig.~\ref{bench}. Similarly, the best
agreement is obtained with $\hbar\tilde\omega=16$ MeV for $^{16}$O.

\section{Comparison with other calculations and experiment} 
\label{comp}

\begin{figure}[hbt]
  \includegraphics[width=0.5\textwidth]{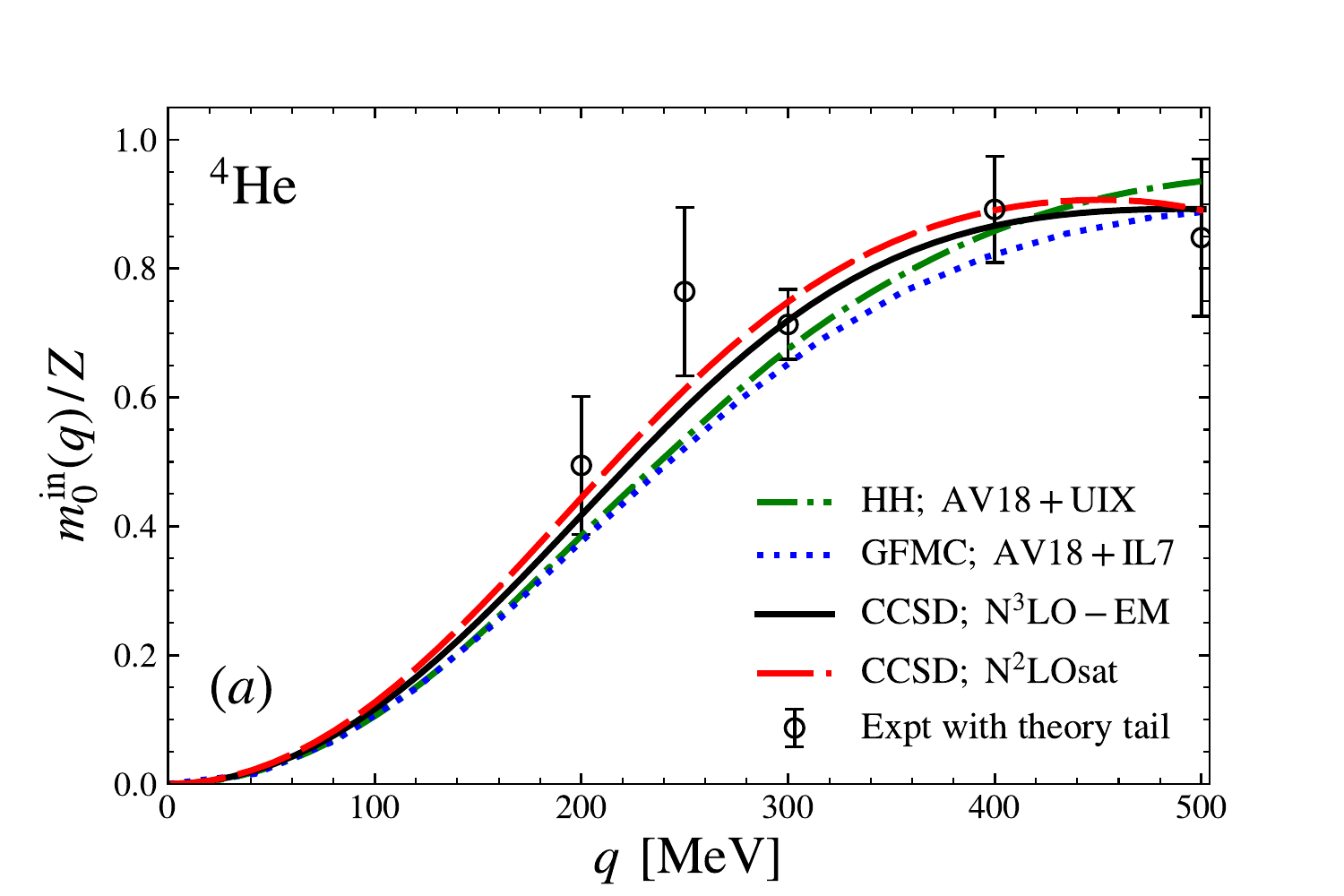}
  \includegraphics[width=0.5\textwidth]{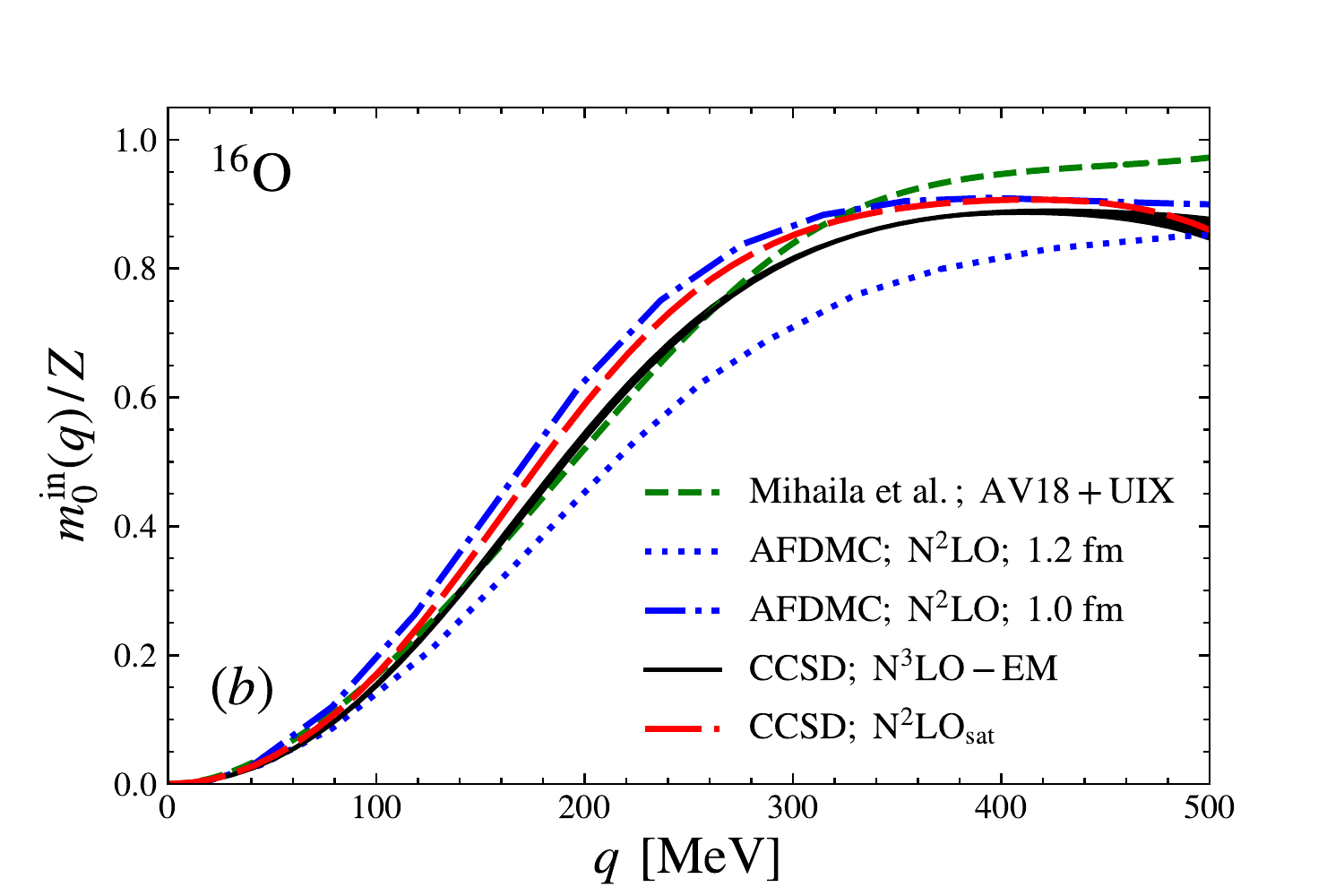}
  \caption{Panel ($a$): CSR for $^4$He in comparison to other calculations of Ref.~\cite{BaccaPRC2009} (AV18+UIX potential)  and
 Refs.~\cite{lovato2013} (AV18+IL7 potential) together with experimental data  from Refs.~\cite{world_data,buki}. Panel ($b$):  CSR for $^{16}$O in comparison to other calculations by Mihaila et al.~\cite{mihaila2000b} (AV18+UIX potential) and by Lonardoni et al.~\cite{Lonardoni} (with local chiral interactions, see text for details). }
  \label{fig-exp}
\end{figure}

Finally, we compare the total CSR calculated from coupled-cluster theory with other available calculations and
experimental data. 
At this stage we include a small effect coming from the nucleon form-factor by rewriting Eq.~\eqref{eq:operator} 
\begin{equation}
\rho(q) = \sum_{j=1}^{A} G_E^p(Q^2) \frac{1+\tau_3}{2} e^{iqz_j}+G_E^n(Q^2) \frac{1-\tau_3}{2} e^{iqz_j}\,,
\end{equation}
where  $(1\pm\tau_3)/2$ are the projectors acting in the isospin space and $Q^2\approx q^2-\omega_{QE}^2$ with the energy transfer $\omega_{QE} = q^2/2m$ corresponding to the quasi-elastic peak, where $m$ is the mass of the nucleon.
We use Kelly's parameterization of $G_E^{n,p}$ form factors~\cite{Kelly:2004hm}, and also account for the Darwin-Foldy relativistic correction by including a factor of $1/[1+Q^2/(4m^2)]$, while neglecting  the smaller spin-orbit contribution.

In Fig.~\ref{fig-exp}, we present the CSR of $^4$He and $^{16}$O for
the N$^3$LO-EM and N$^2$LO$_\mathrm{sat}$ interactions,
respectively. The N$^2$LO$_\mathrm{sat}$ interaction accurately
reproduces the binding energy and charge radius of $^{16}$O, and we therefore
focus on this interaction when comparing with experiment and with
other theoretical approaches that use chiral $NN$ and $3N$ forces.
We noticed a slower convergence with respect to the number of oscillator shells 
in the N$^2$LO$_\mathrm{sat}$ calculations than in the N$^3$LO-EM calculations 
at large values of momentum transfer. The N$^2$LO$_\mathrm{sat}$ results shown in 
Fig.~\ref{fig-exp} is for $\hbar\omega = 20$~MeV and 15 major oscillator shells. 
By extrapolating the observed convergence pattern at smaller model spaces, 
we expect the calculations to be converged at $q<400$~MeV. 
At larger values of $q$, the basis truncation error in CSR for the 
N$^2$LO$_\mathrm{sat}$ interaction 
is expected to be up to a few percentage.

For $^4$He we compare our CCSD calculations against the HH results of
Ref.~\cite{BaccaPRC2009} obtained with the AV18+UIX potential and the
Green's function Monte Carlo (GFMC) results of Refs.~\cite{lovato2013}
obtained with the AV18+IL7 potential.  Although Ref.~\cite{lovato2013}
also included the two-body charge operator, it contributes noticeably
to the CSR of $^4$He only beyond the range of momenta considered in
this study~\cite{Lonardoni:2018nob}.  We see that overall the curves
provide a consistent trend and we note that the dependence on the
implemented Hamiltonian can be interpreted as an overall theoretical
uncertainty.

Regarding $^{4}$He, measurements for the longitudinal response functions at intermediate momentum transfer have been performed in the past and are collected in Ref.~\cite{world_data}, while low momentum data at $q=200$ and $250$ MeV/c are taken from Buki {\it et. al.}~\cite{buki}. Since finite maximal values of the energy transfer $\omega_{max}$ are measured in experimental data, the experimental CSR is obtained as~\cite{tianrui} 
\begin{equation}
m_0^{\mathrm in}(q)=\frac{1}{Z}\int_{\omega^+_{th}}^{\omega_{max}} d\omega\frac{R_L(\omega,q)}{G_E^{P^2}(Q^2)}+m^{\mathrm in}_{0,\text{ tail}}\,,
\end{equation}
where $m^{\mathrm in}_{0,\text{ tail}}$ is taken from the theoretical calculations of $^4$He response functions of Ref.~\cite{th4he}. 
We notice that the experimental trend is well reproduce by all calculations of the CSR.

Our results for the CSR of $^{16}$O are compared with other theoretical calculations in panel ($b$) of Fig.~\ref{fig-exp}. The auxiliary field diffusion Monte Carlo (AFDMC) calculations~\cite{Lonardoni} used local chiral interactions up to next-to-next-to-leading order (N$^2$LO) which were regularized in co-ordinate space with two different cutoff values, $R_0=1.0$~fm and 1.2~fm. The coupled-cluster calculation of Mihaila and Heisenberg~\cite{mihaila2000b} does not include the Darwin-Foldy correction, which would reduce their CSR by as much as 5~\% in the $q\sim500$~MeV region 
and bring it closer to ours.
Further analyses that account for uncertainties due to the interactions and the many-body methods employed are necessary to assess whether the CSR obtained using different theoretical approaches are 
consistent with each other within those uncertainties. Precise experimental data, especially in case of $^{16}$O for which no data exists, can constrain Hamiltonians and many-body methods used in theoretical calculations, and would therefore be very useful.

\section{Conclusions}
\label{concl}
In this paper, we compute the CSR from coupled-cluster theory for
$^4$He and $^{16}$O using $\chi$EFT interactions. For the first time,
we investigate higher momentum transfers regime with these
potentials. Through a benchmark with few-body methods, we show that
the coupled-cluster wave functions retain the relevant correlations
across a broad range of momentum transfer,
even in the CCSD scheme. This is a very promising result because 
accurate treatment of nuclear
correlations in the nuclear electroweak currents in this
range of momentum transfers is important for improving
our understanding of neutrino scattering.

Furthermore, we devise a practical method to remove the spurious
states when working in the lab frame and we show that the method works
 within a precision of a few percent. This is encouraging when
moving to heavier nuclei of interest to neutrino experiments like
$^{40}$Ar where center of mass effects are expected to be smaller. We
provide an analysis of the dominant spuriosities we remove, and find
that these are excited states of a harmonic oscillator with zero
radial excitation and angular excitation equal to $J=1.$

Finally, we compare our results obtained with the N$^3$LO-EM and
\NNLOsat~ potentials with other theoretical results available in the
literature. In the case of $^4$He, we compare also to experimental
data, which we obtained from integrating the longitudinal response
function published in Refs.~\cite{buki,world_data}, and found a nice
agreement. In the case of $^{16}$O no data exist for the CSR, and we
find some sensitivity to the employed Hamiltonian.

This computation of the CSR in coupled-cluster theory constitutes,
indeed, a first important step towards applying this method to
neutrino physics in a broader research program.

\begin{acknowledgments}
  We would like to thank Nir Barnea, Thomas Papenbrock, and Johannes
  Simonis for useful discussions.  This work was supported by the
  Deutsche Forschungsgemeinschaft (DFG) through the Collaborative
  Research Center [The Low-Energy Frontier of the Standard Model (SFB
  1044)], and through the Cluster of Excellence ``Precision Physics,
  Fundamental Interactions, and Structure of Matter" (PRISMA$^+$ EXC
  2118/1) funded by the DFG within the German Excellence Strategy
  (Project ID 39083149), by the Office of Nuclear Physics,
  U.S. Department of Energy, under grants desc0018223 (NUCLEI SciDAC-4
  collaboration) and by the Field Work Proposal ERKBP72 at Oak Ridge
  National Laboratory (ORNL).  Computer time was provided by the
  Innovative and Novel Computational Impact on Theory and Experiment
  (INCITE) program. This research used resources of the Oak Ridge
  Leadership Computing Facility located at ORNL, which is supported by
  the Office of Science of the Department of Energy under Contract
  No. DE-AC05-00OR22725.

\end{acknowledgments}

\FloatBarrier
\bibliographystyle{apsrev}

\end{document}